\documentstyle[12pt]{article}

\def\journal{\topmargin .3in    \oddsidemargin .5in
        \headheight 0pt \headsep 0pt
        \textwidth 5.625in % 1.2 preprint size  %6.5in
\textheight 8.25in % 1.2 preprint size 9in
        \marginparwidth 1.5in
        \parindent 2em
        \parskip .5ex plus .1ex         \jot = 1.5ex}
\journal

\def\ra{\rightarrow}
\begin{document}
\begin{titlepage}
\begin{center}
%\date               \hfill   LBNL-    \\
December 2, 1996       \hfill    LBNL-39621\\
%\hfill hep-ph/
\vskip .5in

{\large \bf Tree-unitary sigma models and their 
application to strong $WW$ scattering}
\footnote
{This work was supported by the Director, Office of Energy
Research, Office of High Energy and Nuclear Physics, Division of High
Energy Physics of the U.S. Department of Energy under Contract
DE-AC03-76SF00098.}

\vskip .5in

Michael S. Chanowitz\footnote{Email: chanowitz@lbl.gov}

\vskip .2in

{\em Theoretical Physics Group\\
     Ernest Orlando Lawrence Berkeley National Laboratory\\
     University of California\\
     Berkeley, California 94720}
\end{center}

\vskip .25in

\begin{abstract}
Sigma models are exhibited which have  tree amplitudes for Goldstone boson 
scattering that satisfy elastic unitarity exactly.  The models have 
imaginary coupling constants and the scalar propagators have poles on 
the imaginary axis in the complex $p^{2}$ plane.  They are 
equivalent to K-matrix models, which are {\it ad hoc} unitarizations 
of low energy theorems for Goldstone boson scattering that
have been used recently to describe strong $WW$ scattering.  
The sigma model formulation of the K-matrix models may be used to 
estimate directly the effect of strong $WW$ scattering on low energy 
radiative corrections.
\end{abstract} 

\end{titlepage}

%THIS PAGE (PAGE ii) CONTAINS THE LBL DISCLAIMER
%TEXT SHOULD BEGIN ON NEXT PAGE (PAGE 1)
\renewcommand{\thepage}{\roman{page}}
\setcounter{page}{2}
\mbox{ }

\vskip 1in

\begin{center}
{\bf Disclaimer}
\end{center}

\vskip .2in

\begin{scriptsize}
\begin{quotation}
This document was prepared as an account of work sponsored by the United
States Government. While this document is believed to contain correct
 information, neither the United States Government nor any agency
thereof, nor The Regents of the University of California, nor any of their
employees, makes any warranty, express or implied, or assumes any legal
liability or responsibility for the accuracy, completeness, or usefulness
of any information, apparatus, product, or process disclosed, or represents
that its use would not infringe privately owned rights.  Reference herein
to any specific commercial products process, or service by its trade name,
trademark, manufacturer, or otherwise, does not necessarily constitute or
imply its endorsement, recommendation, or favoring by the United States
Government or any agency thereof, or The Regents of the University of
California.  The views and opinions of authors expressed herein do not
necessarily state or reflect those of the United States Government or any
agency thereof, or The Regents of the University of California.
\end{quotation}
\end{scriptsize}

\vskip 2in

\begin{center}
\begin{small}
{\it Lawrence Berkeley National Laboratory is an equal opportunity employer.}
\end{small}
\end{center}

\newpage

\renewcommand{\thepage}{\arabic{page}}
\setcounter{page}{1}
%THIS IS PAGE 1 (INSERT TEXT OF REPORT HERE)
%starthere

\noindent {\it \underline {Introduction} }

In perturbation theory unitarity relates terms of different orders and 
cannot be satisfied in tree approximation.  This paper exhibits 
nonstandard sigma models interpreted as effective tree-level theories 
in which the Goldstone boson scattering amplitudes in tree 
approximation do satisfy elastic unitarity exactly.  The models have 
imaginary coupling constants and the sigma propagators have poles on 
the imaginary axis in the complex $p^{2}$ plane.  However, the models 
are not crossing symmetric, a given sigma model only represents a 
specific scattering process, and the discussion is restricted to 
$s$-wave scattering.  Even with these restrictions the models are useful 
in the context in which they were obtained, to represent strong $WW$ 
scattering in a dynamically broken electroweak gauge theory in a 
manner consistent with chiral symmetry and unitarity.  In addition to 
reproducing models of high energy $WW$ scattering, the effective sigma 
model formulation may be used to estimate the contribution of strong 
$WW$ scattering to low energy radiative corrections, which will be 
considered elsewhere.

The tree-unitary sigma models are equivalent to K-matrix unitarization 
of the low energy theorems for Goldstone boson scattering.  The 
K-matrix is an {\it ad hoc} unitarization prescription that has 
been used to construct models of strong $WW$ scattering\cite{kww} at 
high energy colliders.  K-matrix models are suitable for the purpose 
because their partial wave amplitudes are generically ``strong'' 
(i.e., tend to saturate unitarity) while respecting chiral symmetry 
low energy theorems and unitarity.

The equivalence of K-matrix models to tree-unitary Higgs/sigma models 
emerges naturally from a gauge invariant formulation of strong $WW$ 
scattering\cite{noewa1,noewa2} in which models of $s$-wave scattering 
are represented by means of effective ``Higgs boson'' (or sigma) 
propagators.\footnote
{
Because of the equivalence theorem\cite{et} we can refer interchangeably 
to strong scattering of longitudinal $W$'s or Goldstone bosons. 
We also refer interchangeably to Higgs or sigma bosons and to 
Higgs or sigma models.
}
The method is defined by a Feynman diagram algorithm, introduced in 
\cite{noewa2}, which determines the 4-body scattering amplitudes 
involving gauge and/or Goldstone bosons ($WWWW$, $wWWW$, $wwWW$, and 
$wwwW$) from a model of the Landau gauge Goldstone boson scattering 
amplitude ($wwww$).  Tree amplitudes computed from the diagrammatic 
algorithm represent the initial models exactly.  Strong scattering 
models, which are typically formulated in Landau gauge, can then be 
transcribed to unitary gauge or to any generalized renormalizable 
$R_{\xi}$ gauge\cite{fls}.  The initial motivation was to use the 
U-gauge transcription to compute strong $WW$ scattering signals at 
high energy colliders without using the effective $W$ 
approximation\cite{ewa} (EWA), in order to obtain information not 
available from the EWA, such as jet distributions needed for jet tags 
and vetos.  The gauge invariant formulation was checked by direct 
computation\cite{noewa1} and by explicitly verifying BRS 
invariance\cite{noewa2}.

We consider $I=0$ and $I=2$ Goldstone boson scattering channels, which 
both have $s$-wave threshold behavior.  To any model $s$-wave amplitude 
the gauge invariant formulation associates a corresponding effective 
scalar propagator and interaction.  In general, for an 
arbitrarily complicated scattering amplitude, the corresponding scalar 
propagator is arbitrarily complicated and the coupling ``constant'' is 
not constant but is a function of the scattering energy. But  
for K-matrix models the transcription is especially simple: the 
scalar propagators have simple poles, like elementary Higgs scalars, 
and the coupling constants are indeed constant.  However, the pole 
positions are on the negative (positive) imaginary axis for $I=0$  
($I=2$) scattering and the coupling constants are imaginary.

Interpreted naively, the poles correspond to Breit-Wigner resonances 
with decay widths twice as big as their masses.  The imaginary 
coupling constant implies a non-Hermitian Hamiltonian and therefore 
suggests a nonunitary S-matrix, an apparent paradox since the models 
are unitary by construction.  In fact chiral symmetry assures the 
cancellation of the potentially nonunitary terms in the tree 
amplitudes just as it assures the threshold behavior required by the 
low energy theorems. 

The two physical channels with pure $s$-wave threshold behavior are 
$W^{+}_{L}W^{-}_{L}\ra Z_{L}Z_{L}$ and $W^{+}_{L}W^{+}_{L}\ra 
W^{+}_{L}W^{+}_{L}$, where the subscript $L$ denotes longitudinal 
polarization.$^{3}$ The latter is pure $I=2$ while the former is a 
superposition of $I=0$ and $I=2$.  The Higgs boson representation of 
the K-matrix model for $W^{+}_{L}W^{+}_{L}\ra W^{+}_{L}W^{+}_{L}$ 
follows immediately from the transcription defined in reference 
\cite{noewa2}, in which the amplitude is represented by a single 
effective (charge 2) scalar propagator.  
A valid representation of the K-matrix 
model for $W^{+}_{L}W^{-}_{L}\ra Z_{L}Z_{L}$ scattering can also be 
obtained using a single effective propagator, but the propagator does 
not have a single simple pole, the coupling ``constant'' is not 
constant, and the theory does not have a Higgs/sigma model 
structure.  A simple representation is obtained in this case by 
introducing two propagators, corresponding to the two isospin 
components of the $s$-channel amplitude, and the resulting effective 
theory is a two doublet Higgs boson model.

The next section explains the K-matrix prescription and presents the 
K-matrix amplitudes for $W^{+}_{L}W^{+}_{L}\ra W^{+}_{L}W^{+}_{L}$ and 
$W^{+}_{L}W^{-}_{L}\ra Z_{L}Z_{L}$ scattering.  The third section 
describes the effective Higgs boson representation of the K-matrix 
model for $W^{+}_{L}W^{+}_{L}\ra W^{+}_{L}W^{+}_{L}$, while the fourth 
section considers representations of the $W^{+}_{L}W^{-}_{L}\ra 
Z_{L}Z_{L}$ scattering model with one or two effective scalar 
propagators.  The BRS invariance of the effective two doublet model is 
illustrated by explicitly verifying one of the nontrivial BRS 
identities.  The final section contains a brief discussion of the 
results and implications.

\noindent {\it \underline {The K-matrix prescription}}

We consider elastic partial wave unitarity for massless particles 
since we are interested in Goldstone boson scattering or, 
correspondingly, in $W_{L}W_{L}$ scattering at high energy, $E \gg 
m_{W}$, where $m_{W}$ can be neglected.  Strictly speaking there is no 
domain of pure elastic scattering for massless particles, but 
inelastic scattering is strongly suppressed near threshold, and in 
practice $W_{L}W_{L}\ra W_{L}W_{L}W_{L}W_{L}$ is 
negligible relative to $W_{L}W_{L}\ra W_{L}W_{L}$ at the energies of 
interest to us\cite{4w}, between 0.5 and 2 TeV. The unitarity 
constraint on the partial wave 
amplitude $a_{IJ}(s)$ for isospin $I$ and angular momentum $J$ (with 
$s=E^{2}$) is then 
\begin{equation}
{\rm Im}\ a_{IJ} = |a_{IJ}|^{2}.   
\end{equation}

A useful equivalent formulation of equation 1 is 
\begin{equation}
	{\rm Im}\ {1 \over a_{IJ}} = -1. 
\end{equation}	
The K-matrix prescription is defined by choosing an arbitrary real 
function $R_{IJ}(s)$ as the real part of the inverse of $a_{IJ}$,
\begin{equation}
{\rm Re}\ \left({1 \over a_{IJ}}\right)=R_{IJ} 
\end{equation}
and then specifying the complete amplitude $a^{K}_{IJ}$ by 
\begin{equation}
{1 \over a^{K}_{IJ}} = R_{IJ} - i 
\end{equation}
which obviously assures equation 2.

For Goldstone boson scattering, we ensure consistency with the low 
energy theorems\cite{let} that follow from chiral symmetry by 
appropriately choosing the real function $R_{IJ}$,
\begin{equation}
R_{IJ}= {1 \over a_{IJ}^{\rm LET}} 
\end{equation}
where $a_{IJ}^{\rm LET}$ is the low energy theorem amplitude. For the 
the $s$-wave channels the low energy theorem amplitudes are 
\begin{equation}
a_{00}^{\rm LET}= {s \over 16\pi v^{2}} 
\end{equation}
and 
\begin{equation}
a_{20}^{\rm LET}= -{s \over 32\pi v^{2}}.  
\end{equation}

At energies for which the $J=0$ partial 
waves dominate, we have finally the K-matrix models for the $I=0$ and 
$I=2$ channels,
\begin{equation}
{\cal M}_{0}^{K}(s)={s\over v^{2}}\left({1+ia_{00}^{\rm LET}
     \over 1+(a_{00}^{\rm LET})^{2}}\right) 
\end{equation}
and 
\begin{equation}
{\cal M}_{2}^{K}(s)=-{s\over 2v^{2}}\left({1+ia_{20}^{\rm LET}
       \over 1+(a_{20}^{\rm LET})^{2}}\right).
\end{equation}
Including factors of two for states with identical particles the 
isospin decompositions of the physically relevant channels are
\begin{equation}
{\cal M}^{K}(w^{+}w^{-}\ra zz) = 
            {2\over 3}({\cal M}_{0}^{K} - {\cal M}^{K}_{2}) 
\end{equation}
and 
\begin{equation}
{\cal M}^{K}(w^{+}w^{+}\ra w^{+}w^{+}) = 
             2{\cal M}_{2}^{K}. 
\end{equation}

\noindent {\it \underline {Effective Higgs boson model for 
$w^{+}w^{+}\ra w^{+}w^{+}$}}

Because it contains only a single isospin component in the $s$-channel 
the K-matrix model for $w^{+}w^{+}\ra w^{+}w^{+}$ scattering is easily 
expressed as an effective Higgs boson model simply by following the 
algorithm given in \cite{noewa2}.  For an arbitrary model, labeled by 
$X$ and specified by an R-gauge scattering amplitude ${\cal 
M}^{X}_{R}$, the corresponding effective $s$-channel propagator is
\begin{equation}
P^{X}(s) = 
	-\ {v^2 \over s^2}({\cal M}^{X}_{R}- {\cal M}^{\rm LET})
\end{equation}
where ${\cal M}^{\rm LET}$ is the low energy theorem amplitude for the 
relevant channel.  The corresponding Higgs sector coupling constant is
\begin{equation}
\lambda^X(s)= {s\over 2v^2}{{\cal M}^X_R \over {\cal M}^X_R 
                    - {\cal M}_{\rm LET}},
\end{equation}

The vertices that define the Feynman diagram algorithm are given 
in reference \cite{noewa2}.  For $w^{+}w^{+}\ra w^{+}w^{+}$ scattering 
they differ in some instances from the standard model Feynman rules 
because the algorithm imposes an $s$-channel scalar exchange to 
represent interactions that arise from $t$- and $u$-channel exchanges in 
the standard model.\footnote
{
For $w^{+}w^{-}\ra zz$ the algorithm can be represented by an 
effective Lagrangian, but for $w^{+}w^{+}$ 
scattering the model is 
defined only by the Feynman diagram algorithm.   
}
The deviations from the standard model rules are 
specified in table 1 of reference \cite{noewa2}.  In general $P^{X}$ 
may have an arbitrarily complicated form depending on the form of 
${\cal M}^X_R$, and the coupling ``constant'' 
is a function of the scattering energy, 
$\lambda^{X}=\lambda^{X}(s)$.

It is  easy to obtain the effective scalar propagator corresponding 
to the K-matrix model for $w^{+}w^{+}\ra w^{+}w^{+}$. Substituting 
equation 11 and the low energy theorem 
\begin{equation}
{\cal M}^{\rm LET}(w^{+}w^{+}\ra w^{+}w^{+}) = -\ {s\over v^{2}}
\end{equation}
into equation 12 we find the effective propagator has the 
very simple form,
\begin{equation}
P^{K}(w^{+}w^{+}\ra w^{+}w^{+}) = {-1\over s - m_{++}^{2}}
\end{equation}
where 
\begin{equation}
m_{++}^{2}= 32\pi i v^{2}.
\end{equation}
The propagator has a simple pole in the complex $s$ plane, 
though at a peculiar location on the imaginary axis.  
Correspondingly, from equation 13 the coupling constant is in fact 
constant,
\begin{equation}
\lambda_{++}^{K}= 16\pi i
\end{equation}
though with a peculiar imaginary phase.  Notice that the algorithm is 
consistent with the standard model relation 
$m_{++}^{2}=2\lambda_{++}^{K}v^{2}$, which is essential for 
maintaining BRS invariance.  
The negative phase of the propagator arises because, as noted in 
\cite{noewa1,noewa2}, we require an effective $I=2$ 
$s$-channel exchange to represent forces due to $I=0$ $t$- and 
$u$-channel exchanges in the standard model. 

\noindent {\it \underline {Effective Higgs boson model for 
$w^{+}w^{-}\ra zz$}}

A valid BRS invariant representation of the K-matrix model for 
$w^{+}w^{-}\ra zz$ scattering can also be obtained by substituting 
equation 10 and the low energy theorem 
\begin{equation}
{\cal M}^{\rm LET}(w^{+}w^{-}\ra zz) = {s\over v^{2}}
\end{equation}
into equation 12. In this case the effective 
$s$-channel scalar exchange has the same topology as the standard model 
Higgs exchange and with the propagator and coupling constant of 
equations 12 and 13 the Feynman diagram algorithm agrees 
precisely with the standard model Feynman rules.\cite{noewa2} 
The effective propagator is then
\begin{equation}
P^{K}(w^{+}w^{-}\ra zz) = {2\over 3}\left(
          {1\over s -m_{0}^{2}} + 
          {1\over 2}{1\over s -m_{2}^{2}}\right)   
\end{equation}
where
\begin{equation}
m_{0}^{2}= -16\pi i v^{2}   
\end{equation}

\begin{equation}
m_{2}^{2}= +32\pi i v^{2}   
\end{equation}
and the coupling constant is
\begin{equation}
\lambda_{+-}^{K}= {-16\pi i \over 1 +ix}
\end{equation}
where  
\begin{equation}
x = {s\over 16\pi v^{2}}. 
%\eqno (19)
\end{equation} 

The propagator is the sum of two simple scalar poles but the 
coupling ``constant'' is not in fact constant. Although by the 
machinery of reference \cite{noewa2} this is a valid, BRS invariant 
representation of the model, it does not have a simple interpretation 
as a Higgs/sigma model. 

The form of the propagator suggests that we instead consider an 
ansatz with two Higgs scalars.  In particular, consider the two doublet 
model with complex doublets $\Phi_{0}$ and $\Phi_{2}$, corresponding 
to the $I=0$ and $I=2$ components of the $w^{+}w^{-}\ra zz$ amplitude.  
The most general potential with appropriate vacuum is\cite{2h}
\newpage
\begin{eqnarray}
V(\Phi_{0},\Phi_{2})& = &
     \sum_{a=0,2}\lambda_{a}(\Phi_{a}^{\dagger}\Phi_{a}-v_{a}^2)^{2}
     \nonumber \\
   & &     + \lambda_{3}[(\Phi_{0}^{\dagger}\Phi_{0}-v_{0}^{2}) + 
        (\Phi_{2}^{\dagger}\Phi_{2}-v_{2}^{2})]^{2} \nonumber \\
& & +\lambda_{4}[(\Phi_{0}^{\dagger}\Phi_{0})(\Phi_{2}^{\dagger}\Phi_{2})
           -(\Phi_{0}^{\dagger}\Phi_{2})(\Phi_{2}^{\dagger}\Phi_{0})]
          \nonumber  \\
& &  +\lambda_{5}[{\rm Re}(\Phi_{0}^{\dagger}\Phi_{2})-v_{0}v_{2}\ {\rm 
                                       cos}\xi]^{2} \nonumber \\
& &  +\lambda_{6}[{\rm Im}(\Phi_{0}^{\dagger}\Phi_{2})-v_{0}v_{2}\ {\rm 
                                                         sin}\xi]^{2}
\end{eqnarray}
The vacuum expectation values satisfy $v^{2}=v_{0}^{2}+v_{2}^{2}$ with
$m_{W}=gv/2$ and the Goldstone bosons are
\begin{equation}
w^{a}= {\rm cos}\beta\ \phi_{0}^{a} + {\rm sin}\beta\ \phi_{2}^{a}
\end{equation}
where $\phi_{0,2}^{a}$  are the appropriate components 
of the complex doublets  $\Phi_{0,2}$. The angle is determined by the 
ratio of the vev's, tan$\beta = v_{2}/v_{0}$.

The pole positions in equation 19 correspond precisely to 
the $I=0$ and $I=2$ amplitudes,\footnote
{That is, they are the poles 
that would emerge by substituting equations 8 and 9 into equation 12.
}
therefore we want the scalar 
eigenstates to be unmixed, $\alpha=0$ in the conventional notation.
Since we are only constructing an effective tree-level theory to 
replicate the K-matrix amplitude, equation 10, we are free to 
fine-tune the potential shamelessly. We therefore choose 
$\lambda_{3}=\lambda_{5}=0$ so that $H_{0}$ and $H_{2}$ are the 
eigenstates with 
\begin{equation}
m_{a}^{2}=2\lambda_{a}v_{a}^{2}
\end{equation}

To determine the angle $\beta$ we proceed as in \cite{noewa1,noewa2} 
and use the equivalence theorem\cite{et} to determine the U-gauge Higgs 
sector contribution,
\begin{equation}
{\cal M}^{K}_{U,H}= {\cal M}^{K}_{R}-{\cal M}_{\rm gauge\ sector}.
\end{equation}
As always in discussions of strong $WW$ scattering we neglect 
corrections of order $g^{2}$ and $m_{W}/\sqrt{s}$.  In that 
approximation ${\cal M}_{\rm gauge\ sector}\simeq s/v^{2}$ and 
substituting ${\cal M}^{K}_{R}$ from equation 10 we find
\begin{equation}
{\cal M}^{K}_{U,H}= {2\over 3}\ {s^{2}\over v^{2}} \left(
                    {1\over s-m_{0}^{2}}+
                    {1\over 2}\ {1\over s-m_{2}^{2}}\right).
\end{equation}

This is to be compared with the contribution from exchange of the two 
Higgs scalars in the two doublet model,
\begin{equation}
{\cal M}^{2{\rm -doublet}}_{U,H}= \epsilon_{L1}\cdot\epsilon_{L2}
           \ \epsilon_{L3}\cdot\epsilon_{L4} 
           \sum_{a=0,2} \left({g^{2}v_{a}\over 2}\right)^{2}
                    \ {1\over s-m_{a}^{2}}   
\end{equation}
where $\epsilon_{Li}$ are the longitudinal polarization tensors for the 
four gauge bosons. Indices 1 and 2 refer to $W^{\pm}$ and indices 3 
and 4 to the final state $Z$ bosons. 
For simplicity, here and in the discussion of BRS 
invariance below, we assume the gauge group is just $SU(2)_{L}$; I have 
verified that the conclusions are the same for $SU(2)_{L}\times 
U(1)_{Y}$. Approximating $\epsilon_{Li}=p_{i}/m_{W}$ we find that 
equations 28 and 29 are consistent if 
\begin{equation}
{\rm tan}\beta = {1\over 2},
\end{equation}
that is, $v_{0}^{2}=2v^{2}/3$ and $v_{2}^{2}=v^{2}/3$. With the 
masses, equation 20 and 21, this in turn fixes the coupling constants,
\begin{equation}
\lambda_{0}= -12\pi i
\end{equation}
and 
\begin{equation}
\lambda_{2}=+48\pi i.
\end{equation}
It is now straightforward to close the circle by using the 
parameters determined above to verify that the tree amplitude 
${\cal M}(w^{+}w^{-}\ra zz)$ computed from the two doublet model is 
indeed the K-matrix amplitude of equation 10.

We conclude this section by considering one of the BRS identities that 
is nontrivial in the sense that it probes the consistency of gauge and 
Higgs sector interactions,
\begin{equation}
\epsilon_{3\alpha}\epsilon_{4\beta}\left(
k_{1\mu}k_{2\nu}{\cal M}^{\mu\nu\alpha\beta} 
                 +im_W(k_{1\mu}{\cal M}^{\mu\alpha\beta}_{w^{-}}
                       k_{2\nu}{\cal M}^{\nu\alpha\beta}_{w^{+}})
                  -m_W^2{\cal M}^{\alpha\beta}_{w^{+} w^{-}}\right) = 0,
\end{equation}
where subscripts 1,2,3,4 refer to $W^{+},W^{-},Z,Z$ respectively.  The 
subscripts $w^{\pm}$ indicate amplitudes in which gauge boson 
$W^{\pm}$ is replaced by Goldstone boson $w^{\pm}$.  Using the Feynman 
diagram algorithm, which for $W^{+}W^{-}\ra ZZ$ is just the standard 
model Feynman rules, to evaluate the amplitudes in $R_{\xi}$ gauge we 
find after canceling identical terms that the left side of equation 33 
is
\begin{equation}
\delta_{\rm BRS}^2 = {g^2\over 8}\epsilon_3 \cdot \epsilon_4 \left(
              -v^{2}+\sum_{a=0,2}{v_{a}^{2}\over s-m_{a}^{2}}
              (s-2\lambda_{a}v_{a}^{2}) \right)
\end{equation}
which vanishes by equation 26.  Consistency is assured for this and 
the other BRS identities because for $W^{+}W^{-}\ra ZZ$ our 
diagrammatic algorithm is just the usual Feynman rules for the two 
doublet model. (BRS invariance of the algorithm for the $w^{+}w^{+}$ 
channel is less obvious because of departures from the usual 
Feynman rules in that case --- see \cite{noewa2}.)

\noindent {\it \underline {Discussion}}

A similar result to the peculiar pole positions found here was 
obtained in a study of the $I,J=0,0$ channel in the O(2$N$) 
Higgs/sigma model solved to leading order in the $N \ra \infty$ 
limit.\cite{o(n)} Evaluating the solution for $N=2$ (only 33\% worse 
than standard operating procedure for large $N$ QCD) the authors found 
a ``Higgs remnant'' far from the real axis in the fourth quadrant of 
the complex $s$ plane.  In the strong coupling limit the pole position 
tended to $-16\pi i v^{2}/3$, a factor 3 smaller than our K-matrix 
value for $m_{0}^{2}$ (though, as observed by Einhorn\cite{o(n)}, the 
limit is actually outside the domain of validity of the model).

For a heuristic interpretation of the pole positions on the imaginary 
axis in the complex $s$ plane we can consider the Breit-Wigner form, 
\begin{equation}
P_{BW}= {1 \over s - (m - i\Gamma /2)^{2}}
\end{equation}
where $m$ and $\Gamma$ are real. 
For the pole to occur on the imaginary axis, the width must be twice 
the mass, $\Gamma = \pm 2m$.  

The imaginary coupling constants suggest a gross violation of 
unitarity, since a non-Hermitian Hamiltonian implies a nonunitary 
S-matrix, but by construction the tree amplitudes satisfy partial wave 
unitarity exactly.  The explanation is that chiral symmetry protects 
the unitarity of the tree amplitudes just as it assures the threshold 
behavior required by the low energy theorems.  Explicitly, in 
tree approximation the scattering amplitude is the sum of the constant, 
imaginary 4-point contact interaction, $-2\lambda_{a}$, and the 
$s$-channel Higgs/sigma exchange term which contains a canceling 
imaginary constant. Chiral symmetry requires the amplitudes to vanish 
at threshold and therefore enforces the cancellation of the constants  
regardless of their phase.

It is unexpected and interesting that scattering mediated by scalar 
exchanges with poles on the imaginary axis in the $m^{2}$ plane 
corresponds to tree amplitudes that precisely follow the trajectory of 
the Argand circle characterizing exact elastic unitarity.  That 
observation is useful to estimate directly the effect of strong 
$WW$ scattering on low energy radiative corrections.  Most discussions 
of low energy radiative corrections in theories with dynamical 
electroweak symmetry breaking have focused on the effects of specific 
quanta in specific models, for instance, the large oblique corrections 
from techni-quarks or technicolor pseudo-Goldstone 
bosons.\cite{tcradcorr} In the same spirit as the analysis of strong 
$WW$ scattering at high energy experiments, which emphasizes model 
independent aspects of electroweak symmetry breaking by a strong 
force, it would be useful to estimate the effect on low energy 
corrections of just the strong scattering in the $WW$ channels, 
without reference to specific model dependent features.  Such an 
estimate can be made using the Higgs boson representation of the 
K-matrix models and will be considered in a subsequent paper.

\vskip .2in
\noindent Acknowledgements: 
This work was supported by the Director, Office of Energy
Research, Office of High Energy and Nuclear Physics, Division of High
Energy Physics of the U.S. Department of Energy under Contracts
DE-AC03-76SF00098 and DE-AC02-76CHO3000.

\end{document}